\documentclass[twocolumn]{aastex63}

\usepackage{color}
\usepackage{amsmath}

\shorttitle{CID Imaging of Sirius}
\shortauthors{Sawant and Batcheldor}

\begin{document}
\title{CHARGE-INJECTION DEVICE IMAGING OF SIRIUS WITH CONTRAST RATIOS \\ GREATER THAN 1:26 MILLION}
\author[0000-0002-7987-0310]{Sailee M. Sawant}
\affiliation{Department of Aerospace, Physics and Space Sciences, 150 W. University Blvd., Florida Institute of Technology, \\ Melbourne, FL 32901, USA}

\author[0000-0002-8588-5682]{Daniel Batcheldor}
\affiliation{Department of Aerospace, Physics and Space Sciences, 150 W. University Blvd., Florida Institute of Technology, \\ Melbourne, FL 32901, USA}
\affiliation{Southeastern Universities Research Association, Laboratory Support Services and Operations, \\ Kennedy Space Center, FL 32899, USA}

\begin{abstract}
\indent The intrinsic nature of many astronomical objects, such as binary systems, exoplanets, circumstellar and debris disks, and quasar host galaxies, introduces challenging requirements for observational instrumentation and techniques. In each case, we encounter situations where the light from bright sources hampers our ability to detect surrounding fainter targets. To explore all features of such astronomical scenes, we must perform observations at the maximum possible contrast ratios. Charge-injection devices (CIDs) are capable of potentially exceeding contrast ratios of $\log_{10}{(CR)} > 9$ (i.e., 1 part in 1 billion) due to their unique readout architectures and inherent anti-blooming abilities. The on-sky testing of the latest generation of CIDs, the SpectraCAM XDR, has previously demonstrated direct contrast ratios in excess of 1 part in 20 million from sub-optimal ground-based astronomical observations that imposed practical limits on the maximum achievable contrast ratios. Here, we demonstrate the extreme contrast ratio imaging capabilities of the SXDR using observations of Sirius with the 1.0-m Jacobus Kapteyn Telescope, La Palma, Spain. Based on wavelet-based analysis and precise photometric and astrometric calibrations, we report a direct contrast ratio of $\Delta m_r = 18.54$, $\log_{10}{(CR)} = 7.41 \pm 0.08$, or $1$ part in $26$ million. This shows a $29\%$ increase in the achievable contrast ratios compared to the previous results. 
\end{abstract}

\keywords{Astronomy Data Analysis, Astronomical Instrumentation, Astronomical Techniques, Direct Imaging}

\section{Introduction} \label{sec:intro}
\indent Direct imaging of fainter targets in the vicinity of bright sources imposes limitations on the type of contrast observations achievable with ground- and space-based telescopes. The combined effect of small angular separations and large magnitude differences severely impacts the possibility of direct detection of fainter targets \citep{Marois2008, Oppenheimer2009HighContrastOI}. Additionally, the light from bright sources saturates conventional imaging instrumentations, e.g., charge-coupled devices (CCDs) and complementary metal-oxide-semiconductor (CMOS) detectors. Due to their limited full well depth and dynamic range, the directly achievable contrast ratios of these 16-bit detectors are restricted to $\log_{10}{(CR)} < 5$ (i.e., 1 part in 100 thousand). 

\indent Several point-spread function (PSF) suppression techniques are implemented to mitigate bright source signals and subsequently achieve higher contrast ratios (i.e., $ 5 < \log_{10}{(CR)} < 7$). Currently employed techniques, such as coronagraphy \citep[e.g.,][]{SchneiderEtAl.2001}, nulling interferometry \citep[e.g.,][]{BracewellMacPhie1979, Linfield2003}, integral field spectral deconvolution \citep[e.g.,][]{SparksandFord2002}, ground-based angular differential imaging \citep[e.g.,][]{MaroisElAl2006}, and space-based roll subtraction \citep[e.g.,][]{LowranceElAl.2005, SchneiderEtAl2010}, have complex operational requirements that make faint signal detection a challenge. Additionally, these techniques are optimized when stable PSFs are achieved.

\indent  An alternative is to use appropriate instrumentation that can carry out direct extreme contrast ratio (ECR) imaging free from the limitations of currently employed techniques. Charge-injection devices (CIDs) have the intrinsic ability to achieve extreme contrast ratios (ECRs) of $\log_{10}{(CR)} > 9$ (i.e., 1 part in 1 billion) owing to their unique readout architectures and inherent anti-blooming abilities \citep{BhaskaranEtAl2008}. A preliminary study of the Sirius field using the latest generation of commercially available CIDs, the SpectraCAM XDR (SXDR), at the Florida Tech 0.8-m Ortega telescope demonstrated a direct raw contrast ratio of $\Delta m_v = 18.3$, $\log{(CR)} = 7.3$, or 1 part in 20 million at an angular separation of two arcminutes \citep{Batcheldor_2016}. However, the atmospheric conditions in Florida imposed practical limitations on the type of contrast ratios that are otherwise achievable using CIDs. Nonetheless, this result provided a motivation to carry out a direct CID-based ECR imaging of Sirius using the 1.0-m Jacobus Kapteyn Telescope (JKT) at the Roque de Los Muchachos Observatory, La Palma, Spain. 

\indent In this paper, we report the results of the direct ECR imaging of Sirius using the SXDR provided by ThermoFisher Scientific Inc. In $\S$\ref{sec:InstrumentationsandObservations}, we provide a brief introduction to CIDs and discuss the details of our observations. $\S$\ref{sec:reductionmethods} describes the implemented data reduction and analysis. In $\S$\ref{sec:results}, we present the results of the direct ECR imaging of Sirius. In $\S$\ref{sec:discussion}, we discuss these results and outline the future prospects for implementing CIDs as astronomical imaging detectors. $\S$\ref{sec:conclusion} concludes. 

\section{Instrumentations and Observations} \label{sec:InstrumentationsandObservations}
\indent CIDs are an array of X-Y addressable photosensitive metal-oxide-silicon (MOS) capacitors \citep{MICHON1974}. First developed in 1970, the original CID imagers incorporated complex on-chip readout electronics and external (i.e., off-chip) amplifiers that delivered a high read noise of $\sim600\,e^-$ \citep{McCreight:81, McCreightetAl1986}. As a result, CCDs (read noise of $< 10\,e^-$) have been preferred as astronomical detectors. However, modern CIDs are fabricated with pre-amplifier per pixel (PPP) architectures (active pixel technology), random access decoders, and non-destructive readout (NDRO) mechanisms to deliver read noise levels comparable to CCDs \citep{Ninkov1994ChargeID, EidEtAl1995, KimbleEtAl1995, BhaskaranEtAl2008}.

\indent Modern CIDs are also designed to support random access integration (RAI) with extreme dynamic range (XDR). The RAI mode utilizes a random access integrator to identify, monitor, and control the data acquisition on intensely illuminated pixels. A region of interest (ROI) is identified in a short pre-exposure (e.g., 0.1 second). The pixels within the ROI are integrated until an NDRO signal threshold (i.e., $75\%$ of the full well capacity) is reached. The detected signal is readout and stored with a timestamp. The accumulated charge is injected into the underlying substrate, and the pixel operation is repeated. The data is then combined to create a final image with a wider dynamic range.  Since the values of intensely illuminated pixels do not restrict the full well depths of CIDs, the pixels may be integrated, injected, and re-integrated multiple times during a pre-determined exposure.  This allows CIDs to extend the upper limit of the dynamic range, generate 32-bit images, and deliver a potential contrast ratio of $\Delta m_v \sim 24$, $\log_{10}{(CR)} \sim 9.6$, or 1 part in 4 billion  \citep{BhaskaranEtAl2008}. 

\indent Due to their unique pixel architecture, true random accessibility, and NDRO capabilities, CIDs offer several advantages over CCDs: negligible charge-transfer losses, extreme radiation tolerance ($> 5\,$Mrad), large full well capacity, high fill-factor, broad spectral response (156$\,$nm - 1100$\,$nm), and intrinsic anti-blooming abilities \citep{BhaskaranEtAl2008}. Therefore, CIDs are excellent prospects to perform direct ECR imaging with ground- and space-based telescopes. 

\indent The data presented here was acquired using the SXDR installed on the 1.0-m JKT, a facility operated by the Southeastern Association for Research in Astronomy (SARA) at the Roque de Los Muchachos Observatory, La Palma, Spain. The SXDR utilizes a $2048 \times 2048$ $12\,\mu$m pixel Random Access CID820 (RACID) sensor with PPP architecture. The field of view is $10' \times 10'$ with an image scale of $0\farcs30$/pixel. The pixels have a linear (within $2\%$) full well capacity of $268,000\,e^-$. They saturate at $305,000\,e^-$. The peak quantum efficiency at $525\,$nm ($\sim$ V-band) is $48\%$ \citep{BhaskaranEtAl2008}. The detector is hermetically sealed and cooled to $-45.6^\circ$C using an ethylene glycol re-circulation system. At this temperature, the dark current is $5\, e^-/\text{s}$ with a conversion gain of $6.2\,e^-/$ADU. The noise associated with a single read is $44\,e^-$ RMS. This reduces to $5.8\,e^- $ RMS with 128 NDROs for the sampling frequency of 2.1 MHz \citep{Batcheldor_2016}. 

\indent We conducted observations of Sirius on March 8th, 2017, at an airmass of 1.4. We also observed the Landolt standard star field of SA98 670. All images are acquired with the Johnson-Cousins \textit{B, V, R,} and \textit{I} filters. The average DIMM (Differential Image Motion Monitor) seeing was $1\farcs7$ FWHM. The moon illumination was $86\%$.

\section{Data Reduction and Analysis} 
\label{sec:reductionmethods}
\begin{figure}[th!]
\centering
\includegraphics[width = 1.0\linewidth]{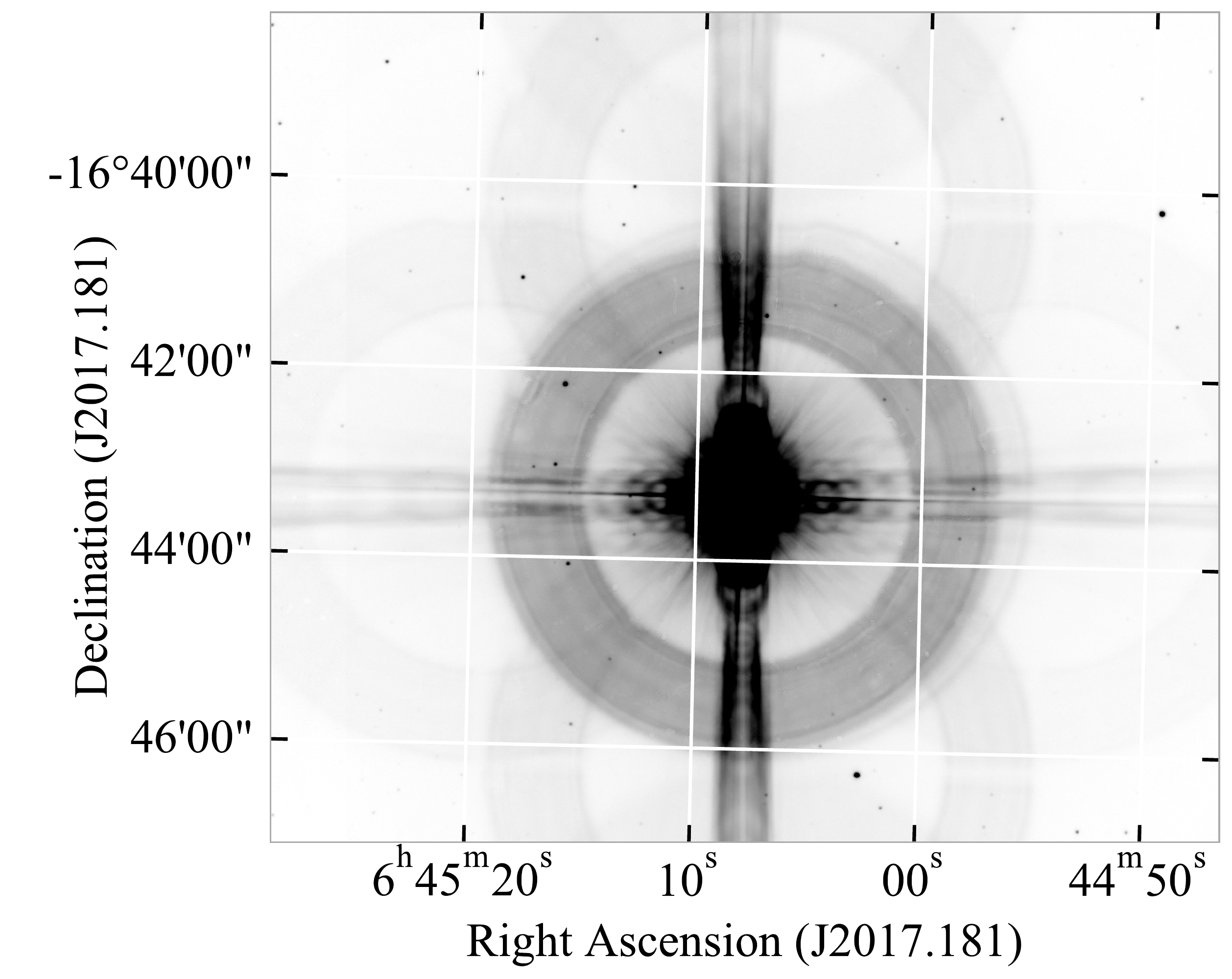}{}
\caption{Pre-processed I-band SXDR image of Sirius with an exposure time of 180 seconds. The field of view is $10' \times 10'$. The signal from Sirius (RA = $06^{\text{h}}45^{\text{m}}08.26^{\text{s}}$, DEC = $-16^\circ43'19\farcs03$) is not saturated. The standard pre-processing routines removed instrumental imperfections; however, it did not suppress the visible artifacts, such as the multiple annular patterns, atmospheric halo, and diffraction spikes.}
\label{fig:preprocessed_SIRIUS_20170308_XDR_I_180}
\end{figure}
Figure \ref{fig:preprocessed_SIRIUS_20170308_XDR_I_180} shows a pre-processed I-band SXDR image of the $10' \times 10'$ field around Sirius at an exposure time of 180 seconds. The signal from Sirius (RA = $06^{\text{h}}45^{\text{m}}08.26^{\text{s}}$, DEC = $-16^\circ43'19\farcs03$) is not saturated. The pre-processing routines successfully removed instrumental imperfections due to dark and flat-field responses; however, an additional post-processing routine was required to suppress the visible artifacts. The internal reflections from the apochromatic field corrector lenses in the JKT produced multiple ghosts (i.e., annular patterns). It is worth noting that such ghosts are absent in the preliminary observations of Sirius \citep[see Figure 4 in][]{Batcheldor_2016}. These artifacts are purely intrinsic to the JKT due to its optical design to acquire a flatter focal plane and a wider field of view \citep{KeelElAt2016}. In addition to this, the atmospheric halo and diffraction spikes dominated the sky background. These effects imposed limitations on the full dynamic range of the image. Therefore, we implemented an additional post-processing routine to mitigate the effects of ghosts, atmospheric halo, and diffraction spikes. 

\subsection{\textnormal{Wavelet Transformation}}
For the cases where the background level varies uniformly, a simple median filter 
is applied to generate a background model, which is then subtracted from pre-processed images \citep[e.g.,][]{SExtractor, photutils}. Additionally, appropriate sigma-clipping techniques are used to achieve better accuracy \citep{Starck2002Chapter2}. Diffused halos from bright sources are suppressed using several methods, such as radial-gradient methods \citep[e.g.,][]{Bonnet-Bidaud2000} and isophotal analysis \citep[e.g.,][]{Batcheldor_2016}.   

\indent However, the presence of multiple ghosts in our data (as seen in Figure \ref{fig:preprocessed_SIRIUS_20170308_XDR_I_180}) resulted in a non-uniform background with abrupt spatial variations. The amalgamation of these patterns with faint source signals highlighted an issue of complex hierarchical structures. Therefore, we applied the \`{a}trous wavelet transform to decompose and analyze the SXDR images at different resolution-related scales \citep{Starck2002Chapter1}. 

\indent The \`{a}trous wavelet transform yields a discrete, stationary, isotropic, and shift-invariant transformation, which makes it well-suited for astronomical image and data analysis \citep{Starck2002Chapter1, MertensandAndrei2015}. We followed the algorithm discussed in \citet{Starck2002Chapter2} and applied the \`{a}trous wavelet transform to the SXDR images. This yielded a set of resolution-related views for a given input SXDR image, which is described as follows:
\begin{equation}
    f(x, y) = \sum_{j=1}^{J} w_j(x, y) + c_J(x, y),
\end{equation}
where $f(x, y)$ is a mathematical representation of the input SXDR image, $w_j(x, y)$ is a set of resolution-related views of the image, called wavelet scales, and $c_J$ is the last smoothed array. The set $w = \mathcal{W}f(x, y) = \{w_1(x, y), ..., w_J(x, y), c_J(x, y) \}$ is known as the wavelet transform of the input SXDR image \citep{MertensandAndrei2015}. 

\indent We modeled the noise in the wavelet space to accurately assess the background variation across the input SXDR image. At each resolution scale $j$, we obtained a wavelet coefficient distribution and carried out a statistical significance test to determine whether a given wavelet coefficient $w_j(x, y)$ is due to signal (i.e., significant) or noise (i.e., not significant). We defined our null hypothesis, $\mathcal{H}_0$, as a value that is locally constant across the input SXDR image. We rejected the hypothesis $\mathcal{H}_0$ when the corresponding $p$-value is less than a pre-specified significance level, $\alpha$ \citep{MillerEtAl2001}. Several thresholding techniques exist for multiple hypothesis testing (e.g., $2\sigma$, $3\sigma$, and the Bonferoni methods). However, these techniques fail to control a critical quantity, the false discovery rate (FDR), the fraction of false rejections based on the total number of rejections \citep{MillerEtAl2001}. The FDR method outmatches the standard techniques by successfully controlling this quantity \citep{HopkinselAt2002}. The corresponding FDR ratio is defined as follows:
\begin{equation}
\begin{aligned}
    \textrm{FDR} = \frac{N^{\mathrm{reject}}_\mathrm{null \; true}}{N^{\mathrm{reject}}}
\end{aligned}
\end{equation}
where $N^{\mathrm{reject}}_\mathrm{null \; true}$ is the number of pixels that are truly not significant but declared significant (i.e., false positives) and $N^{\mathrm{reject}}$ is the total number of pixels declared significant. This method ensures that, on average, the FDR is always less than or equal to $\alpha$ \citep{MillerEtAl2001}. 

We selected an $\alpha$ value such that $0 \le \alpha \le 1$. At each scale, we calculated the $p$-values of the $N$ coefficients, sorted from smallest to largest. We defined and evaluated a quantity, $d$, such that:
\begin{equation}
\begin{aligned}
    d = max \left\{ i: P_i < P_D\right\}, \mathrm{\; where \;} P_D = \frac{i\alpha}{c_N N}
\end{aligned}
\end{equation}
where $P_i$ is the $p$-value of the ith ordered coefficients and $c_N$ is a constant, which is equal to 1 when the $p$-values are statistically independent. We rejected all hypotheses whose $p$-values are less or equal to $P_D$. We then selected the FDR threshold as the wavelet coefficient $w(x, y)$ with the largest index $i$ whose $p$-value satisfied the above inequality. In this way, we calculated an adaptive significant threshold at each scale. A detailed description of this procedure is discussed in \citet{MillerEtAl2001}.

\indent We followed the reconstruction algorithm presented in \citet{Starck2002Chapter2} and produced a filtered version of the input SXDR image. To avoid the loss of flux due to thresholding, we improved our filtering algorithm by using the Van Cittert iteration algorithm \citep{starck_murtagh_bijaoui_1998}. We determined the wavelet transform of the filtered SXDR image and ensured that it produced the same significant wavelet coefficients. For each iteration, we calculated the residual signal. We determined the wavelet transform of the residual signal, applied thresholding, and reconstructed the threshold error signal. The residual signal was added back to the filtered SXDR image. We defined our convergence criteria by calculating the estimated relative error and comparing it to a pre-specified tolerance value, $\epsilon = 0.002$. 

\subsection{\textnormal{Faint Source Detection}}
We used the detection program \textit{photutils.detection}, an affiliated package of AstroPy \citep{astropy_collab, photutils}, to identify sources in the reconstructed noise-reduced SXDR images. This program uses the DAOFIND algorithm \citep{Stetson1987} to identify local maxima against a user-specified threshold value. To detect the faintest sources, we applied the same detection algorithm to the first filtered resolution-view of the SXDR images. This is only possible in wavelet space because the large-scale features, such as the visible artifacts seen in Figure \ref{fig:preprocessed_SIRIUS_20170308_XDR_I_180}, are suppressed at the smallest scale revealing the faint small-scale sources \citep{Starck2002Chapter4}. We extracted the centroid coordinates (\textit{x, y}) of the detected sources with sub-pixel accuracy. Furthermore, we constrained our detection criteria to extract sources that have an ellipticity of 1.0 (i.e., point sources). 

\subsection{\textnormal{Photometric and Astrometric Calibrations}}
\indent The process of transforming our instrumental magnitudes to the standard \textit{UBVRI} magnitudes was carried out using the Gaia Early Data Release 3 (Gaia EDR3) catalog data \citep{GAIA_Mission, GaiaEDR3_PhotometricContentAndValidation}.   
We found our photometric accuracy using the B-, V-, R-, and I-band images of the Landolt standard star field of SA98 670. The resulting mean absolute percentage errors (MAPEs) in the transformed $B, V, R,$ and $I$ magnitudes of the standard stars of SA98 670 are $0.167\%, 0.048\%, 0.072\%, \textrm{ and } 0.068\%$, respectively. 

To match the positions of our detected sources with previously known sources, 
we acquired the astrometric coordinates of the Gaia EDR3 cataloged sources within the same field of view. 
We cross-correlated pixel coordinates of the Gaia EDR3 sources with centroids of the detected sources. We compared the cataloged coordinates of the cross-correlated sources with the corresponding calculated coordinates. The resulting mean absolute accuracies in the right ascension (RA) and declination (DEC) are $0\farcs321$ and $0\farcs258$, respectively. 

\section{Results} \label{sec:results}
\begin{figure}[t!]
\centering
\includegraphics[width = 1.0\linewidth]{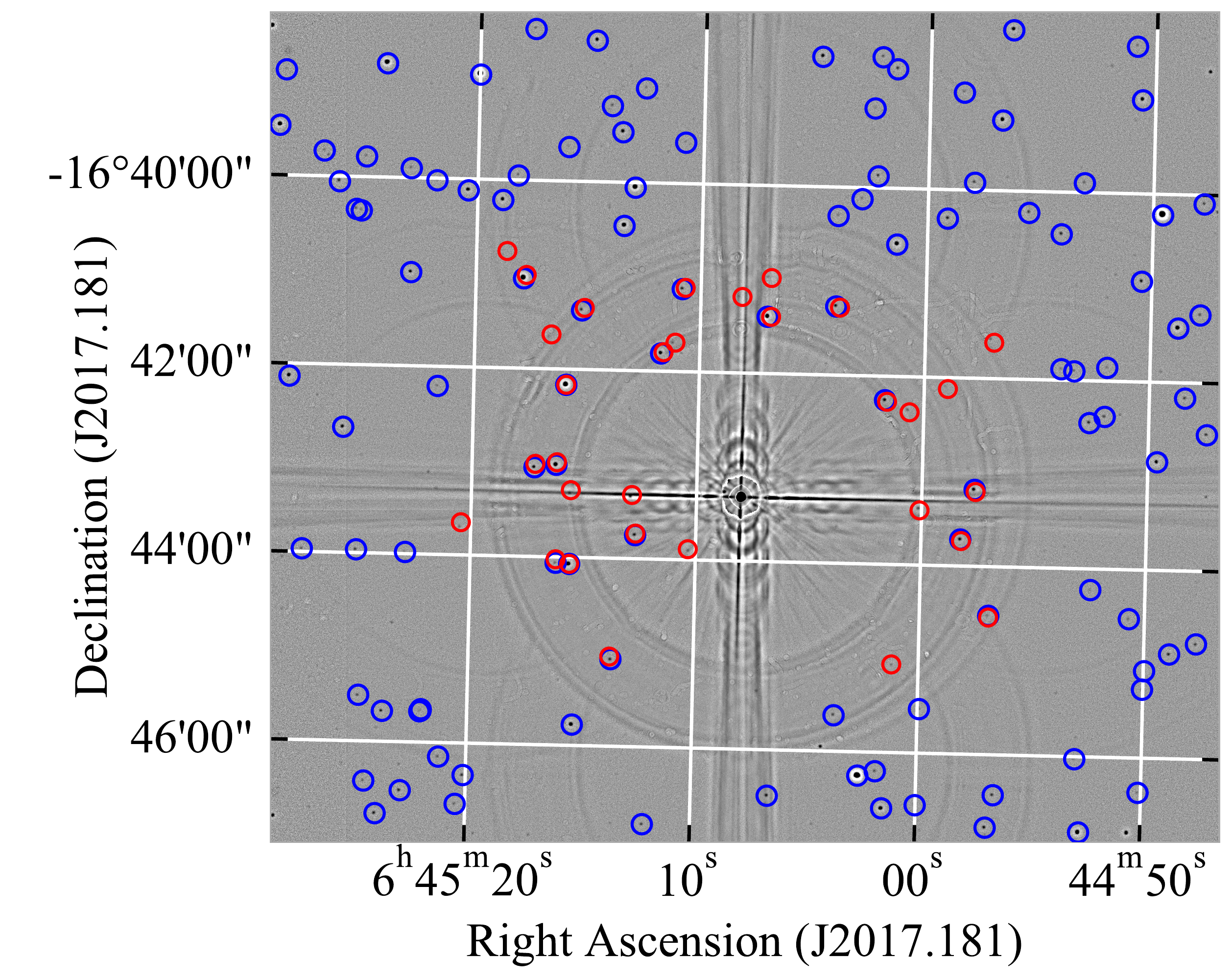}
\caption{Post-processed I-Band SXDR image of the Sirius field with an exposure time of 180 seconds. The SIMBAD and the detected Gaia EDR3 cataloged sources are marked in red and blue, respectively.}
\label{postprocessed_SIRIUS_20170308_XDR_I_180_NEW}
\end{figure} 

\begin{figure*}[t!]
\centering
\includegraphics[width = 1.0\textwidth]{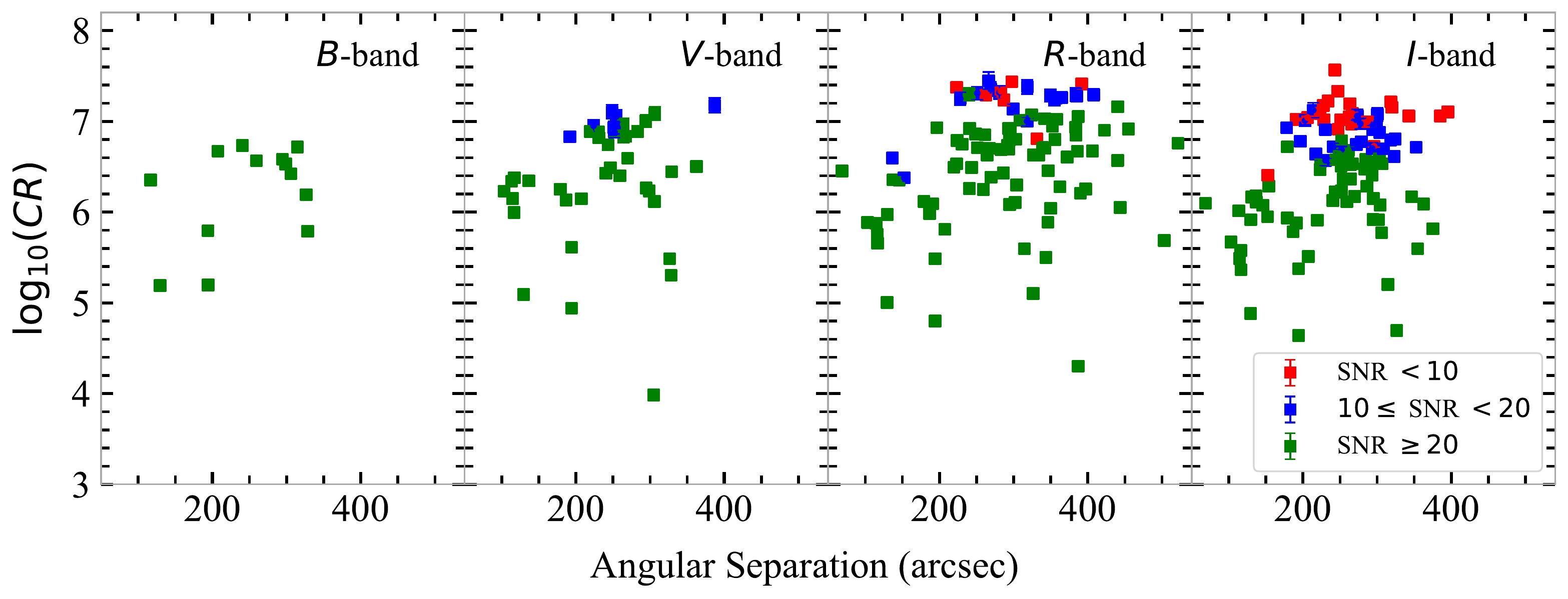}
\caption{Contrast ratios of the detected sources in the B-, V-, R-, and I-band as a function of angular separation (in arcseconds) from Sirius. Detected sources with SNRs $<10$, $10 \leq \textrm{SNRs} < 20$, and SNRs $ \geq 20$ are shown in red, blue, and green, respectively. For sources with SNRs $< 10$, we report their raw contrast ratios. Smaller errors are contained within individual markers.}
\label{fig:2x2PanelContrastRatios}
\end{figure*}

\begin{table*}[]
\begin{center}
\begin{tabular}{cccccc}
\hline
\hline
\multicolumn{1}{p{2cm}}{\centering Filter}
& \multicolumn{1}{p{1cm}}{\centering Source ID} 
& \multicolumn{1}{p{5cm}}{\centering Gaia \\ EDR3 Source ID} 
& \multicolumn{1}{p{3cm}}{\centering Calibrated Magnitude (mag)} 
& \multicolumn{1}{p{3cm}}{\centering $\log_{10}(CR)$}
& \multicolumn{1}{p{1cm}}{\centering SNR} \\ \hline
B & 1 & 2947057338479481600	& 15.216 $\pm$ 0.079 & 6.67 $\pm$ 0.03 & 48.66 \\
V & 2 & 2947046961838568064	& 16.307 $\pm$ 0.189 & 7.11 $\pm$ 0.08 & 15.53 \\
  & 3 & 2947046686960684288	& 15.587 $\pm$ 0.142 & 6.82	$\pm$ 0.06 & 32.46 \\
R & 4 & 2947063042187126400	& 17.077 $\pm$ 0.203 & 7.41 $\pm$ 0.08 & 7.39 \\
  & 5 & 2947062939108083456	& 16.753 $\pm$ 0.166 & 7.29	$\pm$ 0.07 & 11.54\\
  & 6 & 2947051188086163584	& 16.115 $\pm$ 0.113 & 7.03 $\pm$ 0.05 & 24.98 \\
I & 7 & 2947063759445801472	& 16.210 $\pm$ 0.170 & 7.06	$\pm$ 0.07 & 7.73 \\ 
  & 8 & 2947062973476413824	& 15.560 $\pm$ 0.102 & 6.80	$\pm$ 0.04 & 12.36 \\
  & 9 & 2947049882416355712	& 14.837 $\pm$ 0.063 & 6.51 $\pm$ 0.03 & 36.66 \\ \hline
\end{tabular}
\end{center}
\caption{Some of the detected faint Gaia EDR3 cataloged sources in the Sirius field. These sources demonstrate contrast ratios at different SNR levels in each band. No sources with SNRs $< 10 \textrm{ and} < 20$ are detected in the V- and B-band, respectively.}
\label{tab:DetectedGaiaSources}
\end{table*}

\indent Figure \ref{postprocessed_SIRIUS_20170308_XDR_I_180_NEW} shows the post-processed I-band SXDR image of Sirius. The SIMBAD and detected Gaia EDR3 cataloged sources are marked in red and blue, respectively. 

We calculated direct contrast ratios based on the differences between the calibrated apparent magnitudes of the detected sources and the cataloged magnitudes of Sirius (i.e., $\Delta m$). These contrast ratios are usually reported as follows:
\begin{equation}
\log_{10}(CR) = -\frac{\Delta m}{2.5} 
\end{equation}
Figure \ref{fig:2x2PanelContrastRatios} shows direct contrast ratios of the detected Gaia EDR3 cataloged sources as a function of angular separation (in arcseconds) from Sirius. Since the Gaia EDR3 catalog does not include data on Sirius, all contrast ratios and angular separations are calculated with respect to the SIMBAD cataloged data of Sirius \citep{SIMBAD_database}. We acquired the maximum direct contrast ratio of $\Delta m_r = 18.54$, $\log_{10}{(CR)} = 7.41 \pm 0.08$, or $1$ part in $26$ million, a $29\%$ increase from the raw contrast ratio of $\log_{10}{(CR)} = 7.3$ demonstrated from Florida \citep{Batcheldor_2016}. For the cases where the Gaia EDR3 cataloged sources are detected with SNRs $<10$, we calculated their raw contrast ratios (i.e., contrast ratios based on their cataloged magnitudes). These contrast ratios are marked in red. 

\begin{figure}[t!]
\centering
\includegraphics[width = 1.0\linewidth]{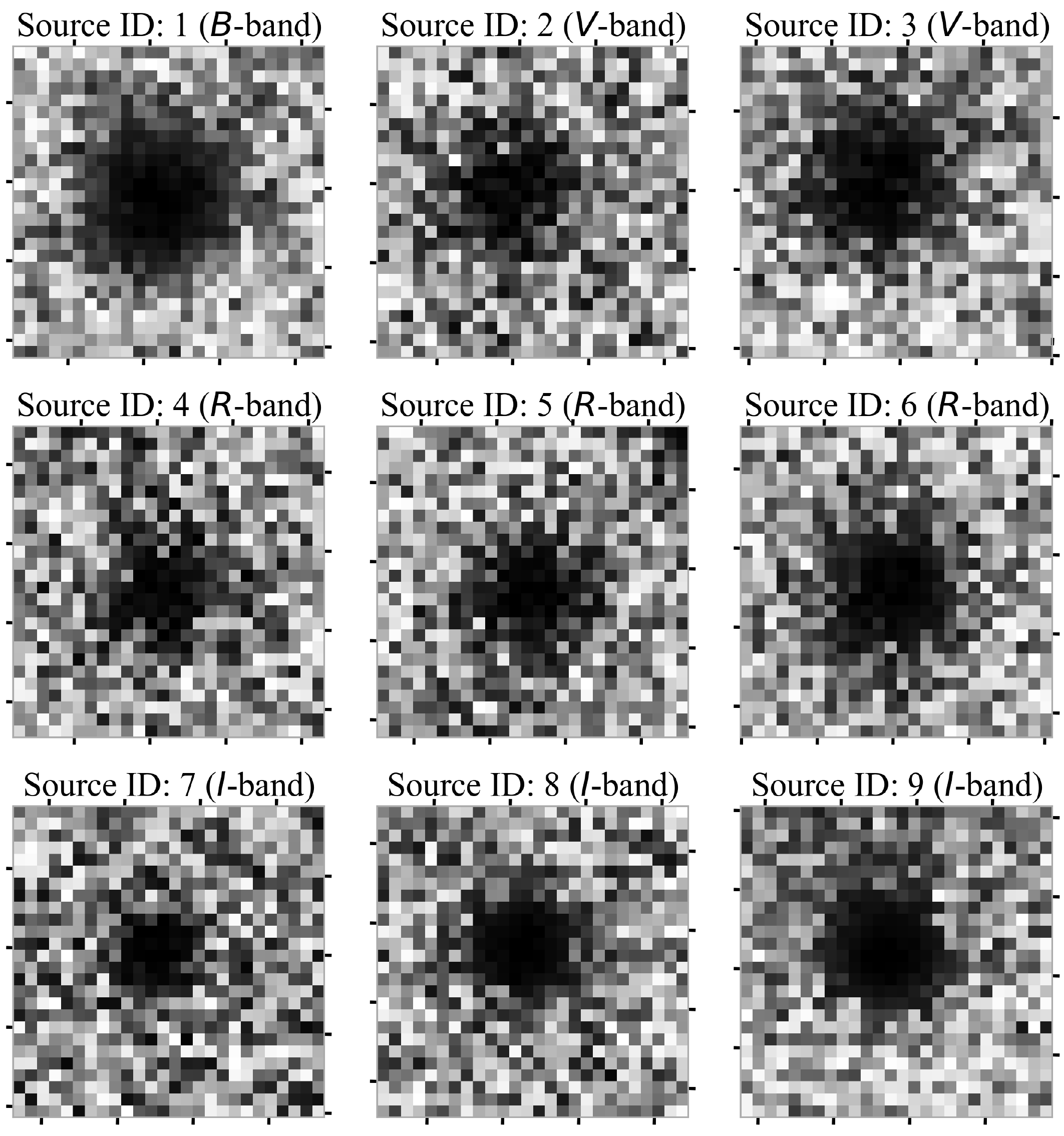}
\caption{$8'' \times 8''$ thumbnails for the detected faint Gaia EDR3 cataloged sources listed in Table \ref{tab:DetectedGaiaSources}.}
\label{fig:3x3PanelThumbnails}
\end{figure}

Table \ref{tab:DetectedGaiaSources} lists some of the detected faint Gaia EDR3 cataloged sources. These sources are detected at different confidence levels based on their SNRs.
The corresponding $8'' \times 8''$ thumbnails are shown in Figure \ref{fig:3x3PanelThumbnails}. No sources with SNRs $<10$ and $<20$ are detected in the V- and B-band, respectively. 

\section{Discussion} \label{sec:discussion}
\indent Due to the extreme brightness of Sirius, previous ground- and space-based observations have prompted the need for appropriate imaging instrumentation and techniques. For instance, \citet{Bonnet-Bidaud1991} and \citet{Bonnet-Bidaud2000} used specially designed coronagraphic devices on ground-based telescopes to look for companions around Sirius. \citet{Bondetal2017} determined several orbital parameters of the Sirius system using an extensive volume of \textit{Hubble Space Telescope} (\textit{HST}) observations and ground-based photographic images. Even with \textit{HST}, significant observing challenges were encountered; no combination of a filter and short exposure time resulted in an unsaturated image of Sirius \citep{Bondetal2017}. However, as we have demonstrated here, it is now possible to acquire an unsaturated signal from Sirius and detect sources at comparable inner working angles (IWAs) using a 1.0-m telescope at an exposure time of 180 seconds without imposing complex operational requirements. 

\begin{figure}[t!]
\centering
\includegraphics[width = 1.0\linewidth]{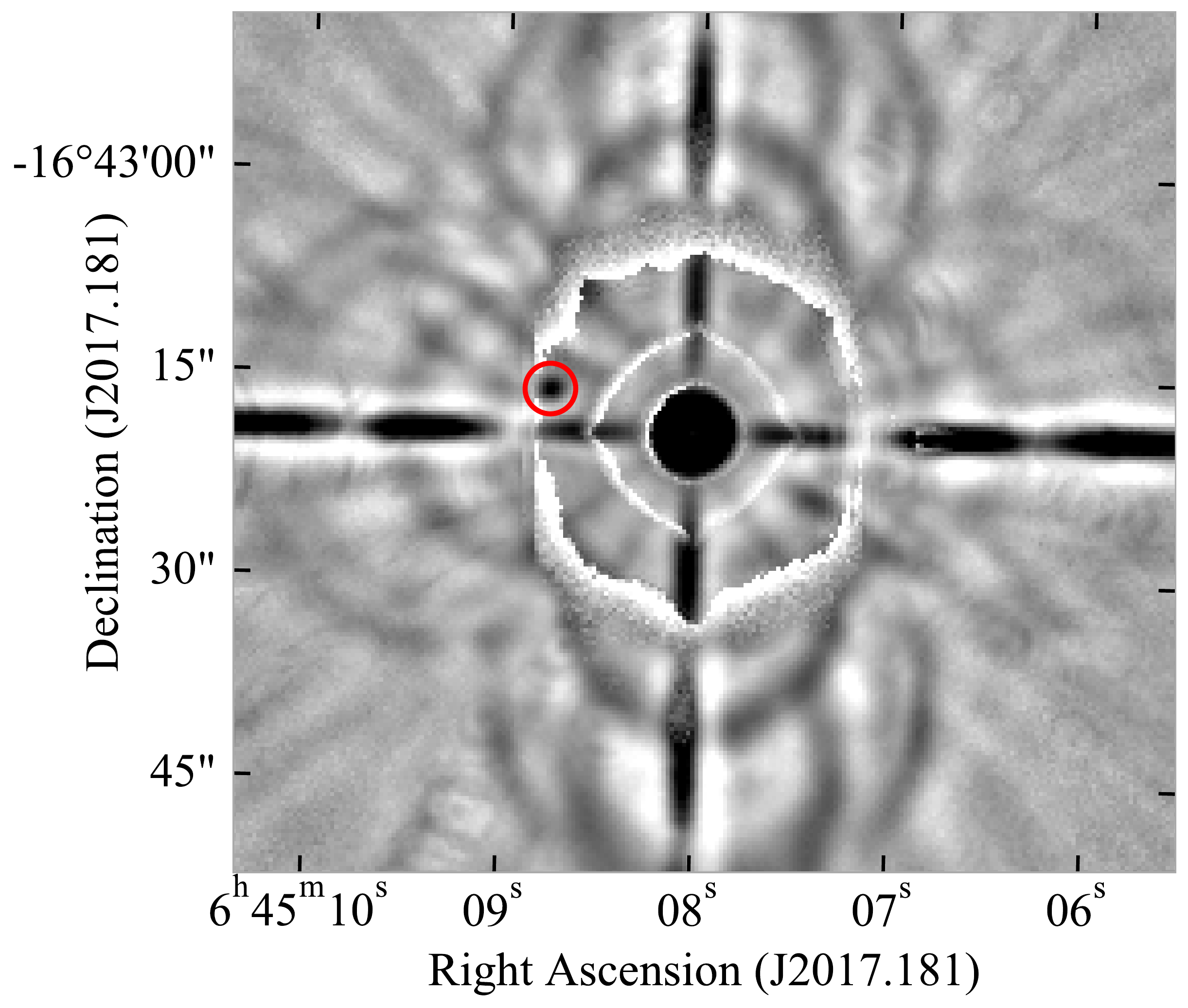}{}
\caption{Central region of the post-processed I-band SXDR image of the Sirius field. Sirius B, annotated in red, is at an angular separation of $8\farcs401$ from Sirius. This is consistent with the proper motion data \citep[e.g.,][]{SIMBAD_database, GAIA_Mission}.}
\label{fig:SiriusB}
\end{figure}

\indent We also detected Sirius B at an angular separation of $8\farcs401$ from Sirius. Figure \ref{fig:SiriusB} shows a zoomed-in view of the post-processed I-Band SXDR image of the field around Sirius. The annotated source is Sirius B. The separation from Sirius is consistent with the proper motion data \citep[e.g.,][]{SIMBAD_database, GAIA_Mission}. This is an improvement in the IWA of the SXDR because Sirius B, despite being $8^{\textrm{th}}$ magnitude, was not detected from Florida \citep{Batcheldor_2016}. 

\indent Within the $5'$ radial field around Sirius, the SIMBAD catalog reports a total of 47 sources out of which [BG91] 1-9 \citep{Bonnet-Bidaud1991} are duplicated with [BCL2000] 1-9 \citep{Bonnet-Bidaud2000}. This is because \citet{Bonnet-Bidaud1991} defined spatial positions as offsets from Sirius, whereas \citet{Bonnet-Bidaud2000} reported the absolute positions. Furthermore, there is a 15-year gap between these observations. We applied corrections to address these source duplicities. Upon analyzing the positions of these sources over the given baseline, we found out that these sources exhibit negligible proper motion. However, the proper motion of Sirius introduced additional systematic offsets to the spatial positions of [BG91]. In addition to this, the photometric and astrometric uncertainties in the [BG91] data are $\pm 0.3$ mag and $\pm 0\farcs7$, respectively \citep{Bonnet-Bidaud1991}.  

Recently, the Gaia EDR3 catalog reports a total of $1181$ sources (without duplicities) within the central $10'$ square field around Sirius. These sources have all-inclusive Gaia photometric and astrometric data with higher precision (see Table 3 in \citet{GaiaCollaborationTable3} for statistical details on uncertainties). Therefore, we used the Gaia EDR3 cataloged sources for our photometric and astrometric calibrations. 

\indent To test the accuracy of our analysis, we applied our reduction methods on a simulated bright star field image. This simulated field consisted of a bright central source along with 99 point sources. We calculated the contrast ratios of these simulated sources with respect to the magnitude of the simulated bright source in a noise-free image. We then added multiple ghost patterns across the image to replicate the non-uniform background. Lastly, we applied our reduction methods and recovered the contrast ratios of the simulated point sources. 
\begin{figure}[t!]
\centering
\includegraphics[width = 1.0\linewidth]{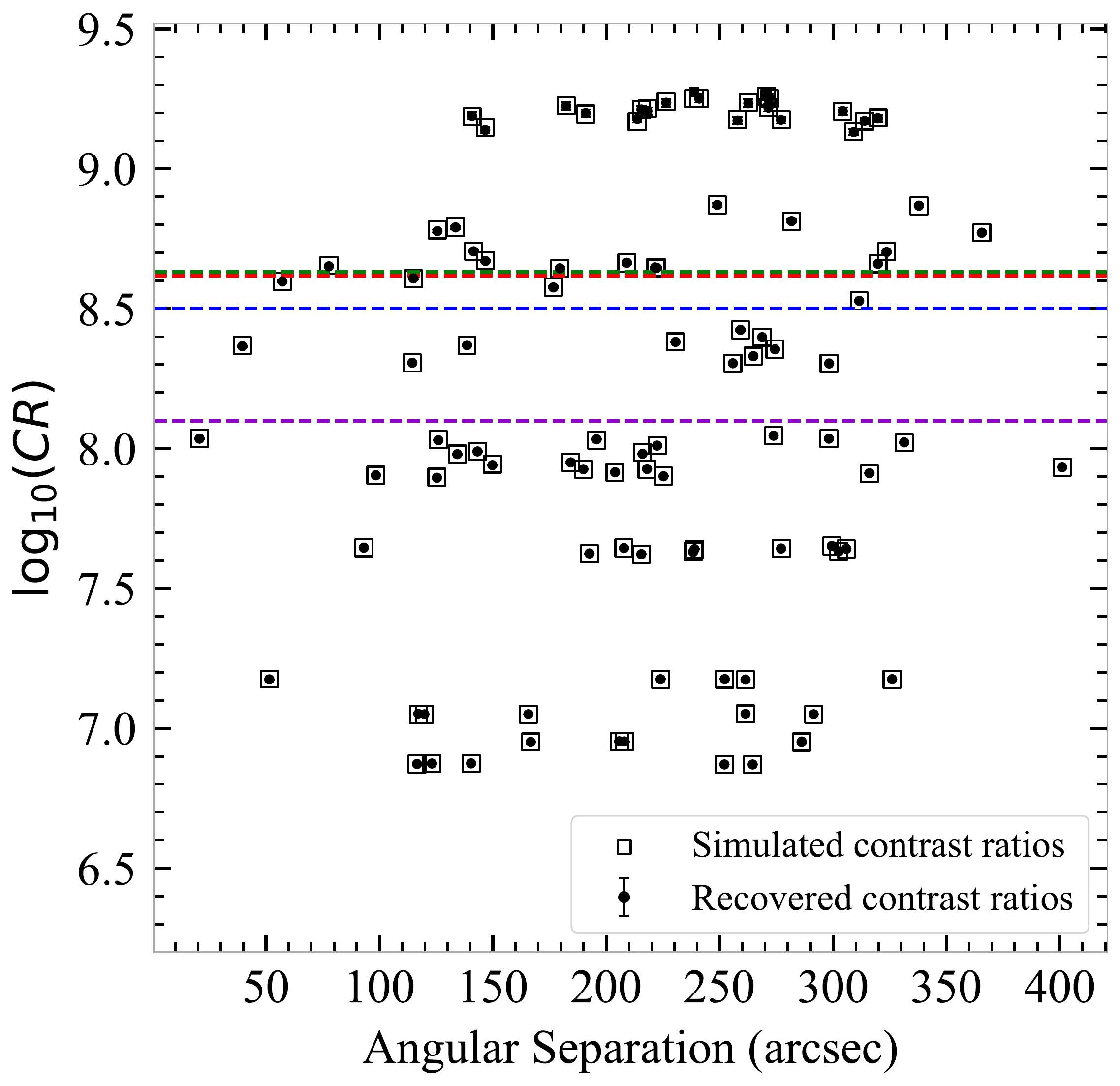}{}
\caption{Plot of $\log_{10}(CR)$ for a set of simulated sources. The contrast ratios and angular separations (in arcseconds) are calculated with respect to a simulated bright source. The unfilled square markers are contrast ratios of pre-defined simulated sources in a noise-free image. The filled circular markers are recovered contrast ratios of simulated sources from a noisy image. The mean error in the recovered contrast ratios is $\pm0.004$. The horizontal dashed colored lines show the sky-limited contrast ratios of $\log_{10}{(CR)} = 8.50\pm0.02, 8.63\pm0.01, 8.62\pm0.01,$ and $8.01\pm0.01$ in the B-band (marked in blue), V-band (marked in green), R-band (marked in red) and I-band (marked in purple), respectively.}
\label{fig:SimulatedBrightStarField}
\end{figure}
Figure \ref{fig:SimulatedBrightStarField} shows a plot of contrast ratios of simulated sources as a function of angular distance (in arcseconds) from the simulated bright source. The unfilled square markers are contrast ratios of pre-defined simulated sources in a noise-free image. The filled circular markers are recovered contrast ratios of these simulated sources from a noisy image. The mean error in the recovered contrast ratios is $\pm 0.004$. The acquired MAPE between the simulated and recovered contrast ratios is $0.016\%$. 

Although we successfully recovered simulated contrast ratios greater than $\log_{10}{(CR)} = 8.0$ (see Figure \ref{fig:SimulatedBrightStarField}), we did not acquire comparable contrast ratios from our Sirius observations. Several instrumental and observational factors imposed practical limitations on the maximum achievable contrast ratios. The optical design of the JKT produced multiple ghost patterns. The moon illumination at the time of data collection increased the overall sky brightness. The resulting non-uniform background and bright sky introduced the limiting magnitudes of $19.793 \pm 0.040, 20.120 \pm 0.028, 20.082 \pm 0.018, \textrm{ and } 18.818 \pm 0.017$ in the B-, V-, R-, and I-band, respectively. The horizontal dashed lines in Figure \ref{fig:SimulatedBrightStarField} represent the corresponding sky-limited contrast ratios of $\log_{10}{(CR)} = 8.50 \pm 0.02, 8.63 \pm 0.01, 8.62 \pm 0.01,$ and $8.01 \pm 0.01$ in the B-band (marked in blue), V-band (marked in green), R-band (marked in red), and I-band (marked in purple), respectively. As clearly seen in Figure \ref{fig:2x2PanelContrastRatios}, we observed fewer sources at shorter wavelengths. This is consistent with the stellar mass function, which is dominated by the latter spectral types. Furthermore, no sources with $\textrm{SNRs} <10$ and $<20$ are detected in the V- and B-band, respectively. Since Sirius is an A1V type star, it is brighter through shorter wavelength filters. Consequently, Rayleigh scattering increased the standard deviation in the background noise level, making it challenging to detect fainter sources.

Nevertheless, we identified several sources with SNRs lower than the minimum threshold. As a result, although visible, these sources did not satisfy our detection requirements. For instance, in the R-band, we identified Gaia EDR3 2947063454512704768, 2947049985495334144, and 2947056616924962688 that have direct raw contrast ratios of $\log_{10}{(CR)} = 7.68\pm0.01, 7.72\pm0.01,$ and $7.73\pm0.01$, respectively. 
Under ideal seeing conditions, it would be possible to detect such sources at $2\sigma$ or $3\sigma$ above the background level. 

CIDs are capable of exceeding contrast ratios of 1 part in 1 billion \citep{BhaskaranEtAl2008}. Even in the presence of practical limitations, we achieved a photometric precision of $\pm0.019, \pm0.049, \pm0.041,$ and $\pm0.044$ mag for $B, V, R,$ and $I$ magnitudes, respectively. As discussed earlier, the contrast ratios acquired from the simulated bright star field data show that the implemented reduction and analysis methods are accurate within a MAPE of $0.016\%$. Furthermore, we found that the mean absolute accuracies in our astrometric calibration are $0\farcs321$ and $0\farcs258$ for RA and DEC, respectively. These methods would deliver even better results for data acquired under ideal observing conditions (e.g., uniform backgrounds and darker skies). Indeed, contrast ratios greater than $\log_{10}{(CR)} = 7.73$ (i.e., 1 part in 54 million) would be easily achievable using CIDs without any operational prerequisites. 

A direct comparison between the number of cataloged sources listed on the SIMBAD and the Gaia EDR3 databases highlights the extent to which such bright star fields remain understudied with the Johnson-Cousins \textit{UBVRI} photometric system. 
Therefore, the potential next step toward practical and cost-effective ground-based ECR imaging demonstration would be to observe different bright star fields from a telescope with minimal internal reflections.
At optimal seeing conditions, it would be possible to detect fainter sources and achieve even higher contrast ratios. To improve IWAs for such observations, it would be effective to incorporate the CID imaging system with relatively simple observing techniques, such as angular differential imaging \citep{MaroisElAl2006}, and software-based PSF modeling techniques.

The presence of concentric circular patterns around Sirius (as seen in Figure \ref{postprocessed_SIRIUS_20170308_XDR_I_180_NEW} and \ref{fig:SiriusB}) suggest the need for some development in the display tools of the SXDR. Since the ROI is assigned during a short pre-exposure, there is a trade-off between the pre-exposure and the atmospheric scintillation timescale. As the SXDR continues to detect photons, the spatial positions of the brightest pixels change within the ROI due to scintillation. To overcome this issue, a future version of the CID technology would incorporate an option to disable the ROI pixels by floating the gate voltages. This would indeed allow CIDs to act as dynamic coronagraphs \citep{Batcheldor_2016}. 

\indent We measured the count rate data of the detected Gaia EDR3 cataloged sources (see Figure \ref{postprocessed_SIRIUS_20170308_XDR_I_180_NEW}, marked in blue) to analyze the ``knockdown" issue. This problem arises due to the imager's inability to detect photons when the pixels in the ROI are being reset. Consequently, some flux is lost when the RAI is combined at the end of exposure time. To characterize the knockdown coefficient, we first calculated the contrast ratios using the cataloged magnitudes of Sirius and the detected Gaia EDR3 cataloged sources (i.e., raw contrast ratios). We then compared these contrast ratios to their corresponding count rate ratios. We found that the mean knockdown coefficients for B-, V-, R-, and I-band images are $0.91, 0.92, 0.90, \textrm{ and } 0.92$, respectively. The observed mean flux ratios are $28\%, 29\%, 24\%, \textrm{ and } 33\%$ lower than their respective true mean flux ratios. 

\indent While knockdown impedes our ability to measure the actual flux from the bright source, this issue is beneficial under certain circumstances. For instance, it helps increase the absolute raw contrast range possible since the light from the bright source is unintentionally suppressed. As discussed in \citet{Batcheldor_2016} and stated above, knockdown can be easily quantified and calibrated using the cataloged magnitudes of known sources in the bright source fields. Therefore, for the cases where absolute photometry is required, this issue can be resolved using the wavelength-dependent knockdown coefficients.

It is important to note that CIDs do not independently address the IWA problem. 
For observations where the target PSFs are not stable or well-known, the IWAs would be large, and detecting nearby fainter sources, such as exoplanets, would be challenging. Since PSFs are most stable in space, CIDs would be most beneficial on space-based telescopes. 
Now that the CIDs are space-qualified to NASA's technology readiness level 8 \citep{Batcheldor2020}, they are excellent prospects to carry out ECR observations with future space-based telescopes. A CID-based high-and extreme-contrast ratio astronomy would most likely help us observe and resolve several ECR scenes, including those with binary systems, exoplanets, circumstellar and debris disks, and quasar host galaxies.

\section{Conclusion} \label{sec:conclusion}
\indent We demonstrated a direct contrast ratio of $\Delta m_r = 18.54$, $\log_{10}{(CR)} = 7.41 \pm 0.08$, or $1$ part in $26$ million using the SXDR on the 1.0-m JKT. We implemented wavelet-based image analysis and acquired filtered SXDR images. We applied photometric and astrometric calibration techniques on the detected GAIA EDR3 sources and calculated contrast ratios with respect to the cataloged magnitude of Sirius. Additionally, we detected and resolved Sirius B at a separation of $8\farcs401$ from Sirius. The optical design of the JKT and high sky brightness imposed practical limitations on the maximum achievable contrast ratios. 
 However, direct contrast ratios greater than $1$ part in $54$ million would be easily achievable with CIDs at optimal observing conditions. \\

\indent 
This research was supported in part by the American Astronomical Society's Small Research Grant Program and the Mount Cuba Astronomical Foundation. We are grateful to the SARA for allocating time on the 1.0-m JKT at the Roque de Los Muchachos Observatory, La Palma, Spain.\\

\bibliography{main}{}

\begin{thebibliography}{}
\expandafter\ifx\csname natexlab\endcsname\relax\def\natexlab#1{#1}\fi
\providecommand{\url}[1]{\href{#1}{#1}}
\providecommand{\dodoi}[1]{doi:~\href{http://doi.org/#1}{\nolinkurl{#1}}}
\providecommand{\doeprint}[1]{\href{http://ascl.net/#1}{\nolinkurl{http://ascl.net/#1}}}
\providecommand{\doarXiv}[1]{\href{https://arxiv.org/abs/#1}{\nolinkurl{https://arxiv.org/abs/#1}}}

\bibitem[{{Astropy Collaboration} {et~al.}(2018){Astropy Collaboration},
  {Price-Whelan}, {Sip{\H{o}}cz}, {G{\"u}nther}, {Lim}, {Crawford}, {Conseil},
  {Shupe}, {Craig}, {Dencheva}, {Ginsburg}, {Vand erPlas}, {Bradley},
  {P{\'e}rez-Su{\'a}rez}, {de Val-Borro}, {Aldcroft}, {Cruz}, {Robitaille},
  {Tollerud}, {Ardelean}, {Babej}, {Bach}, {Bachetti}, {Bakanov}, {Bamford},
  {Barentsen}, {Barmby}, {Baumbach}, {Berry}, {Biscani}, {Boquien}, {Bostroem},
  {Bouma}, {Brammer}, {Bray}, {Breytenbach}, {Buddelmeijer}, {Burke},
  {Calderone}, {Cano Rodr{\'\i}guez}, {Cara}, {Cardoso}, {Cheedella}, {Copin},
  {Corrales}, {Crichton}, {D'Avella}, {Deil}, {Depagne}, {Dietrich}, {Donath},
  {Droettboom}, {Earl}, {Erben}, {Fabbro}, {Ferreira}, {Finethy}, {Fox},
  {Garrison}, {Gibbons}, {Goldstein}, {Gommers}, {Greco}, {Greenfield},
  {Groener}, {Grollier}, {Hagen}, {Hirst}, {Homeier}, {Horton}, {Hosseinzadeh},
  {Hu}, {Hunkeler}, {Ivezi{\'c}}, {Jain}, {Jenness}, {Kanarek}, {Kendrew},
  {Kern}, {Kerzendorf}, {Khvalko}, {King}, {Kirkby}, {Kulkarni}, {Kumar},
  {Lee}, {Lenz}, {Littlefair}, {Ma}, {Macleod}, {Mastropietro}, {McCully},
  {Montagnac}, {Morris}, {Mueller}, {Mumford}, {Muna}, {Murphy}, {Nelson},
  {Nguyen}, {Ninan}, {N{\"o}the}, {Ogaz}, {Oh}, {Parejko}, {Parley}, {Pascual},
  {Patil}, {Patil}, {Plunkett}, {Prochaska}, {Rastogi}, {Reddy Janga},
  {Sabater}, {Sakurikar}, {Seifert}, {Sherbert}, {Sherwood-Taylor}, {Shih},
  {Sick}, {Silbiger}, {Singanamalla}, {Singer}, {Sladen}, {Sooley},
  {Sornarajah}, {Streicher}, {Teuben}, {Thomas}, {Tremblay}, {Turner},
  {Terr{\'o}n}, {van Kerkwijk}, {de la Vega}, {Watkins}, {Weaver}, {Whitmore},
  {Woillez}, {Zabalza}, \& {Astropy Contributors}}]{astropy_collab}
{Astropy Collaboration}, {Price-Whelan}, A.~M., {Sip{\H{o}}cz}, B.~M., {et~al.}
  2018, \aj, 156, 123, \dodoi{10.3847/1538-3881/aabc4f}

\bibitem[{Batcheldor {et~al.}(2016)Batcheldor, Foadi, Bahr, Jenne, Ninkov,
  Bhaskaran, \& Chapman}]{Batcheldor_2016}
Batcheldor, D., Foadi, R., Bahr, C., {et~al.} 2016, Publications of the
  Astronomical Society of the Pacific, 128, 025001,
  \dodoi{10.1088/1538-3873/128/960/025001}

\bibitem[{{Batcheldor} {et~al.}(2020){Batcheldor}, {Sawant}, {Jenne}, {Ninkov},
  {Durrance}, {Bhaskaran}, \& {Chapman}}]{Batcheldor2020}
{Batcheldor}, D., {Sawant}, S., {Jenne}, J., {et~al.} 2020, \pasp, 132, 055001,
  \dodoi{10.1088/1538-3873/ab7a74}

\bibitem[{{Bertin} \& {Arnouts}(1996)}]{SExtractor}
{Bertin}, E., \& {Arnouts}, S. 1996, \aaps, 117, 393,
  \dodoi{10.1051/aas:1996164}

\bibitem[{Bhaskaran {et~al.}(2008)Bhaskaran, Chapman, Pilon, \&
  Vangorden}]{BhaskaranEtAl2008}
Bhaskaran, S., Chapman, T., Pilon, M., \& Vangorden, S. 2008, Proceedings of
  SPIE - The International Society for Optical Engineering, 7055,
  \dodoi{10.1117/12.795235}

\bibitem[{{Bond} {et~al.}(2017){Bond}, {Schaefer}, {Gilliland}, {Holberg},
  {Mason}, {Lindenblad}, {Seitz-McLeese}, {Arnett}, {Demarque}, {Spada},
  {Young}, {Barstow}, {Burleigh}, \& {Gudehus}}]{Bondetal2017}
{Bond}, H.~E., {Schaefer}, G.~H., {Gilliland}, R.~L., {et~al.} 2017, \apj, 840,
  70, \dodoi{10.3847/1538-4357/aa6af8}

\bibitem[{{Bonnet-Bidaud} {et~al.}(2000){Bonnet-Bidaud}, {Colas}, \&
  {Lecacheux}}]{Bonnet-Bidaud2000}
{Bonnet-Bidaud}, J.~M., {Colas}, F., \& {Lecacheux}, J. 2000, \aap, 360, 991.
\newblock \doarXiv{astro-ph/0010032}

\bibitem[{{Bonnet-Bidaud} \& {Gry}(1991)}]{Bonnet-Bidaud1991}
{Bonnet-Bidaud}, J.~M., \& {Gry}, C. 1991, \aap, 252, 193

\bibitem[{{Bracewell} \& {MacPhie}(1979)}]{BracewellMacPhie1979}
{Bracewell}, R.~N., \& {MacPhie}, R.~H. 1979, \icarus, 38, 136,
  \dodoi{10.1016/0019-1035(79)90093-9}

\bibitem[{{Bradley} {et~al.}(2020){Bradley}, {Sipocz}, {Robitaille},
  {Tollerud}, {Deil}, {Vin{\'\i}cius}, {Barbary}, {G{\"u}nther}, {Bostroem},
  {Droettboom}, {Bray}, {Bratholm}, {Pickering}, {Craig}, {Pascual}, {Greco},
  {Donath}, {Kerzendorf}, {Littlefair}, {Barentsen}, {D'Eugenio}, \&
  {Weaver}}]{photutils}
{Bradley}, L., {Sipocz}, B., {Robitaille}, T., {et~al.} 2020,
  astropy/photutils: 1.0.0, 1.0.0,  Zenodo, \dodoi{10.5281/zenodo.4044744}

\bibitem[{Eid(1995)}]{EidEtAl1995}
Eid, S.~I. 1995, in Electronic Imaging

\bibitem[{{Gaia Collaboration} {et~al.}(2016){Gaia Collaboration}, {Prusti},
  {de Bruijne}, {Brown}, {Vallenari}, {Babusiaux}, {Bailer-Jones}, {Bastian},
  {Biermann}, {Evans}, {Eyer}, {Jansen}, {Jordi}, {Klioner}, {Lammers},
  {Lindegren}, {Luri}, {Mignard}, {Milligan}, {Panem}, {Poinsignon},
  {Pourbaix}, {Randich}, {Sarri}, {Sartoretti}, {Siddiqui}, {Soubiran},
  {Valette}, {van Leeuwen}, {Walton}, {Aerts}, {Arenou}, {Cropper}, {Drimmel},
  {H{\o}g}, {Katz}, {Lattanzi}, {O'Mullane}, {Grebel}, {Holland}, {Huc},
  {Passot}, {Bramante}, {Cacciari}, {Casta{\~n}eda}, {Chaoul}, {Cheek}, {De
  Angeli}, {Fabricius}, {Guerra}, {Hern{\'a}ndez}, {Jean-Antoine-Piccolo},
  {Masana}, {Messineo}, {Mowlavi}, {Nienartowicz}, {Ord{\'o}{\~n}ez-Blanco},
  {Panuzzo}, {Portell}, {Richards}, {Riello}, {Seabroke}, {Tanga},
  {Th{\'e}venin}, {Torra}, {Els}, {Gracia-Abril}, {Comoretto},
  {Garcia-Reinaldos}, {Lock}, {Mercier}, {Altmann}, {Andrae}, {Astraatmadja},
  {Bellas-Velidis}, {Benson}, {Berthier}, {Blomme}, {Busso}, {Carry},
  {Cellino}, {Clementini}, {Cowell}, {Creevey}, {Cuypers}, {Davidson}, {De
  Ridder}, {de Torres}, {Delchambre}, {Dell'Oro}, {Ducourant}, {Fr{\'e}mat},
  {Garc{\'\i}a-Torres}, {Gosset}, {Halbwachs}, {Hambly}, {Harrison}, {Hauser},
  {Hestroffer}, {Hodgkin}, {Huckle}, {Hutton}, {Jasniewicz}, {Jordan},
  {Kontizas}, {Korn}, {Lanzafame}, {Manteiga}, {Moitinho}, {Muinonen},
  {Osinde}, {Pancino}, {Pauwels}, {Petit}, {Recio-Blanco}, {Robin}, {Sarro},
  {Siopis}, {Smith}, {Smith}, {Sozzetti}, {Thuillot}, {van Reeven}, {Viala},
  {Abbas}, {Abreu Aramburu}, {Accart}, {Aguado}, {Allan}, {Allasia},
  {Altavilla}, {{\'A}lvarez}, {Alves}, {Anderson}, {Andrei}, {Anglada Varela},
  {Antiche}, {Antoja}, {Ant{\'o}n}, {Arcay}, {Atzei}, {Ayache}, {Bach},
  {Baker}, {Balaguer-N{\'u}{\~n}ez}, {Barache}, {Barata}, {Barbier}, {Barblan},
  {Baroni}, {Barrado y Navascu{\'e}s}, {Barros}, {Barstow}, {Becciani},
  {Bellazzini}, {Bellei}, {Bello Garc{\'\i}a}, {Belokurov}, {Bendjoya},
  {Berihuete}, {Bianchi}, {Bienaym{\'e}}, {Billebaud}, {Blagorodnova},
  {Blanco-Cuaresma}, {Boch}, {Bombrun}, {Borrachero}, {Bouquillon}, {Bourda},
  {Bouy}, {Bragaglia}, {Breddels}, {Brouillet}, {Br{\"u}semeister},
  {Bucciarelli}, {Budnik}, {Burgess}, {Burgon}, {Burlacu}, {Busonero}, {Buzzi},
  {Caffau}, {Cambras}, {Campbell}, {Cancelliere}, {Cantat-Gaudin}, {Carlucci},
  {Carrasco}, {Castellani}, {Charlot}, {Charnas}, {Charvet}, {Chassat},
  {Chiavassa}, {Clotet}, {Cocozza}, {Collins}, {Collins}, {Costigan}, {Crifo},
  {Cross}, {Crosta}, {Crowley}, {Dafonte}, {Damerdji}, {Dapergolas}, {David},
  {David}, {De Cat}, {de Felice}, {de Laverny}, {De Luise}, {De March}, {de
  Martino}, {de Souza}, {Debosscher}, {del Pozo}, {Delbo}, {Delgado},
  {Delgado}, {di Marco}, {Di Matteo}, {Diakite}, {Distefano}, {Dolding}, {Dos
  Anjos}, {Drazinos}, {Dur{\'a}n}, {Dzigan}, {Ecale}, {Edvardsson}, {Enke},
  {Erdmann}, {Escolar}, {Espina}, {Evans}, {Eynard Bontemps}, {Fabre},
  {Fabrizio}, {Faigler}, {Falc{\~a}o}, {Farr{\`a}s Casas}, {Faye}, {Federici},
  {Fedorets}, {Fern{\'a}ndez-Hern{\'a}ndez}, {Fernique}, {Fienga}, {Figueras},
  {Filippi}, {Findeisen}, {Fonti}, {Fouesneau}, {Fraile}, {Fraser}, {Fuchs},
  {Furnell}, {Gai}, {Galleti}, {Galluccio}, {Garabato}, {Garc{\'\i}a-Sedano},
  {Gar{\'e}}, {Garofalo}, {Garralda}, {Gavras}, {Gerssen}, {Geyer}, {Gilmore},
  {Girona}, {Giuffrida}, {Gomes}, {Gonz{\'a}lez-Marcos},
  {Gonz{\'a}lez-N{\'u}{\~n}ez}, {Gonz{\'a}lez-Vidal}, {Granvik}, {Guerrier},
  {Guillout}, {Guiraud}, {G{\'u}rpide}, {Guti{\'e}rrez-S{\'a}nchez}, {Guy},
  {Haigron}, {Hatzidimitriou}, {Haywood}, {Heiter}, {Helmi}, {Hobbs},
  {Hofmann}, {Holl}, {Holland}, {Hunt}, {Hypki}, {Icardi}, {Irwin}, {Jevardat
  de Fombelle}, {Jofr{\'e}}, {Jonker}, {Jorissen}, {Julbe}, {Karampelas},
  {Kochoska}, {Kohley}, {Kolenberg}, {Kontizas}, {Koposov}, {Kordopatis},
  {Koubsky}, {Kowalczyk}, {Krone-Martins}, {Kudryashova}, {Kull}, {Bachchan},
  {Lacoste-Seris}, {Lanza}, {Lavigne}, {Le Poncin-Lafitte}, {Lebreton},
  {Lebzelter}, {Leccia}, {Leclerc}, {Lecoeur-Taibi}, {Lemaitre}, {Lenhardt},
  {Leroux}, {Liao}, {Licata}, {Lindstr{\o}m}, {Lister}, {Livanou}, {Lobel},
  {L{\"o}ffler}, {L{\'o}pez}, {Lopez-Lozano}, {Lorenz}, {Loureiro},
  {MacDonald}, {Magalh{\~a}es Fernandes}, {Managau}, {Mann}, {Mantelet},
  {Marchal}, {Marchant}, {Marconi}, {Marie}, {Marinoni}, {Marrese},
  {Marschalk{\'o}}, {Marshall}, {Mart{\'\i}n-Fleitas}, {Martino}, {Mary},
  {Matijevi{\v{c}}}, {Mazeh}, {McMillan}, {Messina}, {Mestre}, {Michalik},
  {Millar}, {Miranda}, {Molina}, {Molinaro}, {Molinaro}, {Moln{\'a}r},
  {Moniez}, {Montegriffo}, {Monteiro}, {Mor}, {Mora}, {Morbidelli}, {Morel},
  {Morgenthaler}, {Morley}, {Morris}, {Mulone}, {Muraveva}, {Musella},
  {Narbonne}, {Nelemans}, {Nicastro}, {Noval}, {Ord{\'e}novic},
  {Ordieres-Mer{\'e}}, {Osborne}, {Pagani}, {Pagano}, {Pailler}, {Palacin},
  {Palaversa}, {Parsons}, {Paulsen}, {Pecoraro}, {Pedrosa}, {Pentik{\"a}inen},
  {Pereira}, {Pichon}, {Piersimoni}, {Pineau}, {Plachy}, {Plum}, {Poujoulet},
  {Pr{\v{s}}a}, {Pulone}, {Ragaini}, {Rago}, {Rambaux}, {Ramos-Lerate},
  {Ranalli}, {Rauw}, {Read}, {Regibo}, {Renk}, {Reyl{\'e}}, {Ribeiro},
  {Rimoldini}, {Ripepi}, {Riva}, {Rixon}, {Roelens}, {Romero-G{\'o}mez},
  {Rowell}, {Royer}, {Rudolph}, {Ruiz-Dern}, {Sadowski}, {Sagrist{\`a}
  Sell{\'e}s}, {Sahlmann}, {Salgado}, {Salguero}, {Sarasso}, {Savietto},
  {Schnorhk}, {Schultheis}, {Sciacca}, {Segol}, {Segovia}, {Segransan},
  {Serpell}, {Shih}, {Smareglia}, {Smart}, {Smith}, {Solano}, {Solitro},
  {Sordo}, {Soria Nieto}, {Souchay}, {Spagna}, {Spoto}, {Stampa}, {Steele},
  {Steidelm{\"u}ller}, {Stephenson}, {Stoev}, {Suess}, {S{\"u}veges}, {Surdej},
  {Szabados}, {Szegedi-Elek}, {Tapiador}, {Taris}, {Tauran}, {Taylor},
  {Teixeira}, {Terrett}, {Tingley}, {Trager}, {Turon}, {Ulla}, {Utrilla},
  {Valentini}, {van Elteren}, {Van Hemelryck}, {van Leeuwen}, {Varadi},
  {Vecchiato}, {Veljanoski}, {Via}, {Vicente}, {Vogt}, {Voss}, {Votruba},
  {Voutsinas}, {Walmsley}, {Weiler}, {Weingrill}, {Werner}, {Wevers},
  {Whitehead}, {Wyrzykowski}, {Yoldas}, {{\v{Z}}erjal}, {Zucker}, {Zurbach},
  {Zwitter}, {Alecu}, {Allen}, {Allende Prieto}, {Amorim},
  {Anglada-Escud{\'e}}, {Arsenijevic}, {Azaz}, {Balm}, {Beck}, {Bernstein},
  {Bigot}, {Bijaoui}, {Blasco}, {Bonfigli}, {Bono}, {Boudreault}, {Bressan},
  {Brown}, {Brunet}, {Bunclark}, {Buonanno}, {Butkevich}, {Carret}, {Carrion},
  {Chemin}, {Ch{\'e}reau}, {Corcione}, {Darmigny}, {de Boer}, {de Teodoro}, {de
  Zeeuw}, {Delle Luche}, {Domingues}, {Dubath}, {Fodor}, {Fr{\'e}zouls},
  {Fries}, {Fustes}, {Fyfe}, {Gallardo}, {Gallegos}, {Gardiol}, {Gebran},
  {Gomboc}, {G{\'o}mez}, {Grux}, {Gueguen}, {Heyrovsky}, {Hoar}, {Iannicola},
  {Isasi Parache}, {Janotto}, {Joliet}, {Jonckheere}, {Keil}, {Kim},
  {Klagyivik}, {Klar}, {Knude}, {Kochukhov}, {Kolka}, {Kos}, {Kutka}, {Lainey},
  {LeBouquin}, {Liu}, {Loreggia}, {Makarov}, {Marseille}, {Martayan},
  {Martinez-Rubi}, {Massart}, {Meynadier}, {Mignot}, {Munari}, {Nguyen},
  {Nordlander}, {Ocvirk}, {O'Flaherty}, {Olias Sanz}, {Ortiz}, {Osorio},
  {Oszkiewicz}, {Ouzounis}, {Palmer}, {Park}, {Pasquato}, {Peltzer}, {Peralta},
  {P{\'e}turaud}, {Pieniluoma}, {Pigozzi}, {Poels}, {Prat}, {Prod'homme},
  {Raison}, {Rebordao}, {Risquez}, {Rocca-Volmerange}, {Rosen}, {Ruiz-Fuertes},
  {Russo}, {Sembay}, {Serraller Vizcaino}, {Short}, {Siebert}, {Silva},
  {Sinachopoulos}, {Slezak}, {Soffel}, {Sosnowska}, {Strai{\v{z}}ys}, {ter
  Linden}, {Terrell}, {Theil}, {Tiede}, {Troisi}, {Tsalmantza}, {Tur},
  {Vaccari}, {Vachier}, {Valles}, {Van Hamme}, {Veltz}, {Virtanen}, {Wallut},
  {Wichmann}, {Wilkinson}, {Ziaeepour}, \& {Zschocke}}]{GAIA_Mission}
{Gaia Collaboration}, {Prusti}, T., {de Bruijne}, J.~H.~J., {et~al.} 2016,
  \aap, 595, A1, \dodoi{10.1051/0004-6361/201629272}

\bibitem[{{Gaia Collaboration} {et~al.}(2021){Gaia Collaboration}, {Brown},
  {Vallenari}, {Prusti}, {de Bruijne}, {Babusiaux}, {Biermann}, {Creevey},
  {Evans}, {Eyer}, {Hutton}, {Jansen}, {Jordi}, {Klioner}, {Lammers},
  {Lindegren}, {Luri}, {Mignard}, {Panem}, {Pourbaix}, {Randich}, {Sartoretti},
  {Soubiran}, {Walton}, {Arenou}, {Bailer-Jones}, {Bastian}, {Cropper},
  {Drimmel}, {Katz}, {Lattanzi}, {van Leeuwen}, {Bakker}, {Cacciari},
  {Casta{\~n}eda}, {De Angeli}, {Ducourant}, {Fabricius}, {Fouesneau},
  {Fr{\'e}mat}, {Guerra}, {Guerrier}, {Guiraud}, {Jean-Antoine Piccolo},
  {Masana}, {Messineo}, {Mowlavi}, {Nicolas}, {Nienartowicz}, {Pailler},
  {Panuzzo}, {Riclet}, {Roux}, {Seabroke}, {Sordo}, {Tanga}, {Th{\'e}venin},
  {Gracia-Abril}, {Portell}, {Teyssier}, {Altmann}, {Andrae}, {Bellas-Velidis},
  {Benson}, {Berthier}, {Blomme}, {Brugaletta}, {Burgess}, {Busso}, {Carry},
  {Cellino}, {Cheek}, {Clementini}, {Damerdji}, {Davidson}, {Delchambre},
  {Dell'Oro}, {Fern{\'a}ndez-Hern{\'a}ndez}, {Galluccio}, {Garc{\'\i}a-Lario},
  {Garcia-Reinaldos}, {Gonz{\'a}lez-N{\'u}{\~n}ez}, {Gosset}, {Haigron},
  {Halbwachs}, {Hambly}, {Harrison}, {Hatzidimitriou}, {Heiter},
  {Hern{\'a}ndez}, {Hestroffer}, {Hodgkin}, {Holl}, {Jan{\ss}en}, {Jevardat de
  Fombelle}, {Jordan}, {Krone-Martins}, {Lanzafame}, {L{\"o}ffler}, {Lorca},
  {Manteiga}, {Marchal}, {Marrese}, {Moitinho}, {Mora}, {Muinonen}, {Osborne},
  {Pancino}, {Pauwels}, {Petit}, {Recio-Blanco}, {Richards}, {Riello},
  {Rimoldini}, {Robin}, {Roegiers}, {Rybizki}, {Sarro}, {Siopis}, {Smith},
  {Sozzetti}, {Ulla}, {Utrilla}, {van Leeuwen}, {van Reeven}, {Abbas}, {Abreu
  Aramburu}, {Accart}, {Aerts}, {Aguado}, {Ajaj}, {Altavilla}, {{\'A}lvarez},
  {{\'A}lvarez Cid-Fuentes}, {Alves}, {Anderson}, {Anglada Varela}, {Antoja},
  {Audard}, {Baines}, {Baker}, {Balaguer-N{\'u}{\~n}ez}, {Balbinot}, {Balog},
  {Barache}, {Barbato}, {Barros}, {Barstow}, {Bartolom{\'e}}, {Bassilana},
  {Bauchet}, {Baudesson-Stella}, {Becciani}, {Bellazzini}, {Bernet}, {Bertone},
  {Bianchi}, {Blanco-Cuaresma}, {Boch}, {Bombrun}, {Bossini}, {Bouquillon},
  {Bragaglia}, {Bramante}, {Breedt}, {Bressan}, {Brouillet}, {Bucciarelli},
  {Burlacu}, {Busonero}, {Butkevich}, {Buzzi}, {Caffau}, {Cancelliere},
  {C{\'a}novas}, {Cantat-Gaudin}, {Carballo}, {Carlucci}, {Carnerero},
  {Carrasco}, {Casamiquela}, {Castellani}, {Castro-Ginard}, {Castro Sampol},
  {Chaoul}, {Charlot}, {Chemin}, {Chiavassa}, {Cioni}, {Comoretto}, {Cooper},
  {Cornez}, {Cowell}, {Crifo}, {Crosta}, {Crowley}, {Dafonte}, {Dapergolas},
  {David}, {David}, {de Laverny}, {De Luise}, {De March}, {De Ridder}, {de
  Souza}, {de Teodoro}, {de Torres}, {del Peloso}, {del Pozo}, {Delbo},
  {Delgado}, {Delgado}, {Delisle}, {Di Matteo}, {Diakite}, {Diener},
  {Distefano}, {Dolding}, {Eappachen}, {Edvardsson}, {Enke}, {Esquej}, {Fabre},
  {Fabrizio}, {Faigler}, {Fedorets}, {Fernique}, {Fienga}, {Figueras},
  {Fouron}, {Fragkoudi}, {Fraile}, {Franke}, {Gai}, {Garabato},
  {Garcia-Gutierrez}, {Garc{\'\i}a-Torres}, {Garofalo}, {Gavras}, {Gerlach},
  {Geyer}, {Giacobbe}, {Gilmore}, {Girona}, {Giuffrida}, {Gomel}, {Gomez},
  {Gonzalez-Santamaria}, {Gonz{\'a}lez-Vidal}, {Granvik},
  {Guti{\'e}rrez-S{\'a}nchez}, {Guy}, {Hauser}, {Haywood}, {Helmi}, {Hidalgo},
  {Hilger}, {H{\l}adczuk}, {Hobbs}, {Holland}, {Huckle}, {Jasniewicz},
  {Jonker}, {Juaristi Campillo}, {Julbe}, {Karbevska}, {Kervella}, {Khanna},
  {Kochoska}, {Kontizas}, {Kordopatis}, {Korn}, {Kostrzewa-Rutkowska},
  {Kruszy{\'n}ska}, {Lambert}, {Lanza}, {Lasne}, {Le Campion}, {Le Fustec},
  {Lebreton}, {Lebzelter}, {Leccia}, {Leclerc}, {Lecoeur-Taibi}, {Liao},
  {Licata}, {Lindstr{\o}m}, {Lister}, {Livanou}, {Lobel}, {Madrero Pardo},
  {Managau}, {Mann}, {Marchant}, {Marconi}, {Marcos Santos}, {Marinoni},
  {Marocco}, {Marshall}, {Martin Polo}, {Mart{\'\i}n-Fleitas}, {Masip},
  {Massari}, {Mastrobuono-Battisti}, {Mazeh}, {McMillan}, {Messina},
  {Michalik}, {Millar}, {Mints}, {Molina}, {Molinaro}, {Moln{\'a}r},
  {Montegriffo}, {Mor}, {Morbidelli}, {Morel}, {Morris}, {Mulone}, {Munoz},
  {Muraveva}, {Murphy}, {Musella}, {Noval}, {Ord{\'e}novic}, {Orr{\`u}},
  {Osinde}, {Pagani}, {Pagano}, {Palaversa}, {Palicio}, {Panahi}, {Pawlak},
  {Pe{\~n}alosa Esteller}, {Penttil{\"a}}, {Piersimoni}, {Pineau}, {Plachy},
  {Plum}, {Poggio}, {Poretti}, {Poujoulet}, {Pr{\v{s}}a}, {Pulone}, {Racero},
  {Ragaini}, {Rainer}, {Raiteri}, {Rambaux}, {Ramos}, {Ramos-Lerate}, {Re
  Fiorentin}, {Regibo}, {Reyl{\'e}}, {Ripepi}, {Riva}, {Rixon}, {Robichon},
  {Robin}, {Roelens}, {Rohrbasser}, {Romero-G{\'o}mez}, {Rowell}, {Royer},
  {Rybicki}, {Sadowski}, {Sagrist{\`a} Sell{\'e}s}, {Sahlmann}, {Salgado},
  {Salguero}, {Samaras}, {Sanchez Gimenez}, {Sanna}, {Santove{\~n}a},
  {Sarasso}, {Schultheis}, {Sciacca}, {Segol}, {Segovia}, {S{\'e}gransan},
  {Semeux}, {Shahaf}, {Siddiqui}, {Siebert}, {Siltala}, {Slezak}, {Smart},
  {Solano}, {Solitro}, {Souami}, {Souchay}, {Spagna}, {Spoto}, {Steele},
  {Steidelm{\"u}ller}, {Stephenson}, {S{\"u}veges}, {Szabados}, {Szegedi-Elek},
  {Taris}, {Tauran}, {Taylor}, {Teixeira}, {Thuillot}, {Tonello}, {Torra},
  {Torra}, {Turon}, {Unger}, {Vaillant}, {van Dillen}, {Vanel}, {Vecchiato},
  {Viala}, {Vicente}, {Voutsinas}, {Weiler}, {Wevers}, {Wyrzykowski}, {Yoldas},
  {Yvard}, {Zhao}, {Zorec}, {Zucker}, {Zurbach}, \&
  {Zwitter}}]{GaiaCollaborationTable3}
{Gaia Collaboration}, {Brown}, A.~G.~A., {Vallenari}, A., {et~al.} 2021, \aap,
  649, A1, \dodoi{10.1051/0004-6361/202039657}

\bibitem[{{Hopkins} {et~al.}(2002){Hopkins}, {Miller}, {Connolly}, {Genovese},
  {Nichol}, \& {Wasserman}}]{HopkinselAt2002}
{Hopkins}, A.~M., {Miller}, C.~J., {Connolly}, A.~J., {et~al.} 2002, \aj, 123,
  1086, \dodoi{10.1086/338316}

\bibitem[{Keel {et~al.}(2016)Keel, Oswalt, Mack, Henson, Hillwig, Batcheldor,
  Berrington, Pree, Hartmann, Leake, Licandro, Murphy, Webb, \&
  Wood}]{KeelElAt2016}
Keel, W.~C., Oswalt, T., Mack, P., {et~al.} 2016, Publications of the
  Astronomical Society of the Pacific, 129, 015002,
  \dodoi{10.1088/1538-3873/129/971/015002}

\bibitem[{Kimble {et~al.}(1995)Kimble, Chen, Haas, Norton, Payne, Carbone, \&
  Corba}]{KimbleEtAl1995}
Kimble, R.~A., Chen, P.~C., Haas, J.~P., {et~al.} 1995, in EUV, X-Ray, and
  Gamma-Ray Instrumentation for Astronomy VI, ed. O.~H.~W. Siegmund \& J.~V.
  Vallerga, Vol. 2518, International Society for Optics and Photonics (SPIE),
  397 -- 409, \dodoi{10.1117/12.218392}

\bibitem[{Linfield(2003)}]{Linfield2003}
Linfield, R.~P. 2003, in Interferometry in Space, ed. M.~Shao, Vol. 4852,
  International Society for Optics and Photonics (SPIE), 443 -- 450,
  \dodoi{10.1117/12.460862}

\bibitem[{{Lowrance} {et~al.}(2005){Lowrance}, {Becklin}, {Schneider},
  {Kirkpatrick}, {Weinberger}, {Zuckerman}, {Dumas}, {Beuzit}, {Plait},
  {Malumuth}, {Heap}, {Terrile}, \& {Hines}}]{LowranceElAl.2005}
{Lowrance}, P.~J., {Becklin}, E.~E., {Schneider}, G., {et~al.} 2005, \aj, 130,
  1845, \dodoi{10.1086/432839}

\bibitem[{{Marois} {et~al.}(2006){Marois}, {Lafreni{\`e}re}, {Doyon},
  {Macintosh}, \& {Nadeau}}]{MaroisElAl2006}
{Marois}, C., {Lafreni{\`e}re}, D., {Doyon}, R., {Macintosh}, B., \& {Nadeau},
  D. 2006, \apj, 641, 556, \dodoi{10.1086/500401}

\bibitem[{Marois {et~al.}(2008)Marois, Macintosh, Barman, Zuckerman, Song,
  Patience, Lafreniere, \& Doyon}]{Marois2008}
Marois, C., Macintosh, B., Barman, T., {et~al.} 2008, Science, 322,
  1348–1352, \dodoi{10.1126/science.1166585}

\bibitem[{McCreight \& Goebel(1981)}]{McCreight:81}
McCreight, C.~R., \& Goebel, J.~H. 1981, Appl. Opt., 20, 3189,
  \dodoi{10.1364/AO.20.003189}

\bibitem[{McCreight {et~al.}(1986)McCreight, McKelvey, Goebel, Anderson, \&
  Lee}]{McCreightetAl1986}
McCreight, C.~R., McKelvey, M.~E., Goebel, J.~H., Anderson, G.~M., \& Lee,
  J.~H. 1986, in Infrared Detectors, Sensors, and Focal Plane Arrays, ed.
  H.~Nakamura, Vol. 0686, International Society for Optics and Photonics
  (SPIE), 66 -- 75, \dodoi{10.1117/12.936526}

\bibitem[{{Mertens} \& {Lobanov}(2015)}]{MertensandAndrei2015}
{Mertens}, F., \& {Lobanov}, A. 2015, \aap, 574, A67,
  \dodoi{10.1051/0004-6361/201424566}

\bibitem[{{Michon, G.}(1974)}]{MICHON1974}
{Michon, G.} 1974, uS Patent 3,786,263

\bibitem[{{Miller} {et~al.}(2001){Miller}, {Genovese}, {Nichol}, {Wasserman},
  {Connolly}, {Reichart}, {Hopkins}, {Schneider}, \& {Moore}}]{MillerEtAl2001}
{Miller}, C.~J., {Genovese}, C., {Nichol}, R.~C., {et~al.} 2001, \aj, 122,
  3492, \dodoi{10.1086/324109}

\bibitem[{Ninkov {et~al.}(1994)Ninkov, Tang, Backer, Easton, \&
  Carbone}]{Ninkov1994ChargeID}
Ninkov, Z., Tang, C., Backer, B.~S., Easton, R.~L., \& Carbone, J.~M. 1994, in
  Astronomical Telescopes and Instrumentation

\bibitem[{Oppenheimer \& Hinkley(2009)}]{Oppenheimer2009HighContrastOI}
Oppenheimer, B.~R., \& Hinkley, S. 2009, Annual Review of Astronomy and
  Astrophysics, 47, 253, \dodoi{10.1146/annurev-astro-082708-101717}

\bibitem[{{Riello} {et~al.}(2021){Riello}, {De Angeli}, {Evans}, {Montegriffo},
  {Carrasco}, {Busso}, {Palaversa}, {Burgess}, {Diener}, {Davidson}, {Rowell},
  {Fabricius}, {Jordi}, {Bellazzini}, {Pancino}, {Harrison}, {Cacciari}, {van
  Leeuwen}, {Hambly}, {Hodgkin}, {Osborne}, {Altavilla}, {Barstow}, {Brown},
  {Castellani}, {Cowell}, {De Luise}, {Gilmore}, {Giuffrida}, {Hidalgo},
  {Holland}, {Marinoni}, {Pagani}, {Piersimoni}, {Pulone}, {Ragaini}, {Rainer},
  {Richards}, {Sanna}, {Walton}, {Weiler}, \&
  {Yoldas}}]{GaiaEDR3_PhotometricContentAndValidation}
{Riello}, M., {De Angeli}, F., {Evans}, D.~W., {et~al.} 2021, \aap, 649, A3,
  \dodoi{10.1051/0004-6361/202039587}

\bibitem[{{Schneider} {et~al.}(2001){Schneider}, {Becklin}, {Smith},
  {Weinberger}, {Silverstone}, \& {Hines}}]{SchneiderEtAl.2001}
{Schneider}, G., {Becklin}, E.~E., {Smith}, B.~A., {et~al.} 2001, \aj, 121,
  525, \dodoi{10.1086/318050}

\bibitem[{{Schneider} {et~al.}(2010){Schneider}, {Silverstone}, {Stobie},
  {Rhee}, \& {Hines}}]{SchneiderEtAl2010}
{Schneider}, G., {Silverstone}, M.~D., {Stobie}, E., {Rhee}, J.~H., \& {Hines},
  D.~C. 2010, in Hubble after SM4. Preparing JWST, 15

\bibitem[{{Sparks} \& {Ford}(2002)}]{SparksandFord2002}
{Sparks}, W.~B., \& {Ford}, H.~C. 2002, \apj, 578, 543, \dodoi{10.1086/342401}

\bibitem[{Starck \& Murtagh(2002{\natexlab{a}})}]{Starck2002Chapter2}
Starck, J.-L., \& Murtagh, F. 2002{\natexlab{a}}, Filtering (Berlin,
  Heidelberg: Springer Berlin Heidelberg), 27--58,
  \dodoi{10.1007/978-3-662-04906-8_2}

\bibitem[{Starck \& Murtagh(2002{\natexlab{b}})}]{Starck2002Chapter1}
---. 2002{\natexlab{b}}, Introduction to Applications and Methods (Berlin,
  Heidelberg: Springer Berlin Heidelberg), 1--25,
  \dodoi{10.1007/978-3-662-04906-8_1}

\bibitem[{Starck \& Murtagh(2002{\natexlab{c}})}]{Starck2002Chapter4}
---. 2002{\natexlab{c}}, Detection (Berlin, Heidelberg: Springer Berlin
  Heidelberg), 93--113, \dodoi{10.1007/978-3-662-04906-8_4}

\bibitem[{Starck {et~al.}(1998)Starck, Murtagh, \&
  Bijaoui}]{starck_murtagh_bijaoui_1998}
Starck, J.-L., Murtagh, F.~D., \& Bijaoui, A. 1998, Image Processing and Data
  Analysis: The Multiscale Approach (Cambridge University Press),
  \dodoi{10.1017/CBO9780511564352}

\bibitem[{{Stetson}(1987)}]{Stetson1987}
{Stetson}, P.~B. 1987, \pasp, 99, 191, \dodoi{10.1086/131977}

\bibitem[{{Wenger} {et~al.}(2000){Wenger}, {Ochsenbein}, {Egret}, {Dubois},
  {Bonnarel}, {Borde}, {Genova}, {Jasniewicz}, {Lalo{\"e}}, {Lesteven}, \&
  {Monier}}]{SIMBAD_database}
{Wenger}, M., {Ochsenbein}, F., {Egret}, D., {et~al.} 2000, \aaps, 143, 9,
  \dodoi{10.1051/aas:2000332}

\end{thebibliography}
\bibliographystyle{aasjournal}

\end{document}